# Research on the background correction method in x-ray phase contrast imaging with Talbot-Lau interferometer [*]

Wang Sheng-Hao (王圣浩)[1], Zhang Can (张灿)[1], Margie P. Olbinado[2], Atsushi Momose[2], Gao Kun (高昆)[1], Wang Zhi-Li (王志立)[1], Hu Ren-Fang (胡仁芳)[1], Han Hua-Jie (韩华杰)[1], Yang Meng (杨萌)[1], Zhang Kai (张凯)[3], Zhu Pei-Ping (朱佩平)[3] and Wu Zi-Yu (吴自玉)[1, 3; ‡]

[1] National Synchrotron Radiation Laboratory, University of Science and Technology of China, Hefei 230027, China.

[2] Institute of Multidisciplinary Research for Advanced Materials, Tohoku University, 2-1-1 Katahira, Aoba-ku, Sendai, Miyagi 980-8577, Japan.

[3] Institute of High Energy Physics, Chinese Academy of Sciences, Beijing 100049, China.

**Abstract:** X-ray Talbot-Lau interferometer has been used widely to conduct phase contrast imaging with a conventional low-brilliance x-ray source. Typically, in this technique, background correction has to be performed in order to obtain the pure signal of the sample under inspection. In this study, we reported on a research on the background correction strategies within this technique, especially we introduced a new phase unwrapping solution for one conventional background correction method, the key point of this solution is changing the initial phase of each pixel by a novel cyclic shift operation on the raw images collected in phase stepping scan. Experimental result and numerical analysis showed that the new phase unwrapping algorithm could successfully subtract contribution of the system's background without error. Moreover, a potential advantage of this phase unwrapping strategy is that its effective phase measuring range could be tuned flexibly in some degree for example to be $(-\pi+3, \pi+3]$, thus it would find usage in certain case because the effective measuring range of the currently widely used background correction method is fixed to be $(-\pi, \pi]$.

[*] Supported by the Major State Basic Research Development Program of China (Grant No. 2012CB825800), the Science Fund for Creative Research Groups, China (Grant No. 11321503), the National Natural Science Foundation of China (Grant Nos. 11205189, 11475170, and 11205157), and the Japan-Asia Youth Exchange program in Science (*SAKURA Exchange Program in Science*) administered by the Japan Science and Technology Agency.

‡ Email: wuzy@ustc.edu.cn

# 1. Introduction

X-ray phase-contrast imaging, which uses phase shift as the imaging signal, can provide remarkably improved contrast over conventional absorption-based imaging for weakly absorbing samples, such as biological soft tissues and fibre composites. [1-3] In the last 50 years, several x-ray phase-contrast imaging methods have been put forward, [4-13] however none of them has so far found wide applications in medical or industrial areas, where typically the use of a laboratory x-ray source is highly required. The demonstration of a Talbot-Lau interferometer in the hard x-ray region with a conventional low-brilliance x-ray source can be considered as a breakthrough in x-ray phase-contrast imaging, [14-16] because it showed that phase-contrast x-ray imaging can be successfully conducted with a low-brilliance x-ray tube, thus overcoming the problems that impaired a wider use of phase-contrast in x-ray radiography and tomography, and many potential applications of this technique in biomedical imaging have been studied. [17-21]

During x-ray phase contrast imaging with Talbot-Lau interferometer, typically a background correction process has to be performed, in the hope of obtaining the pure phase signal of the sample. [13, 22] Therefore in the experiment, usually one phase stepping scan is carried out with sample, while another scan needs to be performed without sample in the beam path, and then in the data post processing, based on these two sets of raw data, two background correction methods written respectively as formulae (1.1) and (1.2) are usually used to retrieve the refraction image of the sample.

$$\varphi(x,y) = \frac{d}{2\pi \times z_T} \times \arg \left\{ \frac{\sum_{k=0}^{M-1}[I_{k+1}^s(x,y) \times \exp(-i2\pi \frac{k}{M})]}{\sum_{k=0}^{M-1}[I_{k+1}^b(x,y) \times \exp(-i2\pi \frac{k}{M})]} \right\}, \quad (1.1)$$

$$\varphi(x,y) = \frac{d}{2\pi \times z_T} \times \left\{ \arg \sum_{k=0}^{M-1}[I_{k+1}^s(x,y) \times \exp(-i2\pi \frac{k}{M})] - \arg \sum_{k=0}^{M-1}[I_{k+1}^b(x,y) \times \exp(-i2\pi \frac{k}{M})] \right\}, \quad (1.2)$$

where $z_T$ is the distance between gratings G1 and G2 (when the sample is placed between gratings G0 and G1), $d$ stands for the period of the grating G2. $I_{k+1}^{s}(x,y)$ and $I_{k+1}^{b}(x,y)$ represent the gray values of the pixel $(x,y)$ at the $(k+1)^{th}$ step of the phase stepping scan with and without sample, respectively. $M$ is the number of steps during the scan in one period of grating G2.

In the following, we name formula (1.1) as method arg(S/B) and formula (1.2) as method arg(S)-arg(B). These two methods look equivalent in mathematics, however there are some differences between them when performing background correction in phase contrast imaging with an x-ray Talbot-Lau interferometer. The effective phase measuring range of method arg(S)-arg(B) is (-2π, 2π], but phase wrapping phenomenon sometimes takes place even when the phase shift introduced by the sample is very small, this is the reason why method arg(S/B) gets used more widely in the community than method arg(S)-arg(B). However the effective phase measuring range of method arg(S/B) is fixed to be (-π, π], the expected phase would be wrapped if it exceeds this range, and complicated phase unwrapping algorithm has to be carried out in order to get the real phase value. [23-27] Combining advantages of these two methods in some degree and meanwhile avoiding weaknesses of them would thus be interesting and meaningful.

The following sections will be arranged like this, firstly we will display layout of the x-ray Talbot-Lau interferometer and acquisition of the raw data, and then comparison is made between methods arg(S)-arg(B) and arg(S/B) when dealing with a set of data. After that we analyze the origin of phase wrapping in method arg(S)-arg(B), and provide a conventional phase unwrapping solution. Then the theory of circle shift operation on the raw images is demonstrated, based on which we will introduce the new phase unwrapping strategy, performance of the new phase wrapping solution together with the conventional one would then be evaluated in comparison with methods arg(S)-arg(B) and arg(S/B). Finally, potential advantage of the new phase unwrapping solution would be discussed and two identical but more convenient implementation means of the new solution in PC will be provided.

## 2. Materials and methods

### 2.1 Experimental setup

The x-ray phase-contrast imaging was carried out with a Talbot-Lau interferometer located at the Institute of Multidisciplinary Research for Advanced Materials, Tohoku University, Japan. Fig. 1 is the schematic of this x-ray interferometer, it is mainly made up of an x-ray tube, three micro-structured gratings, and an x-ray detector, which are assembled on multi-dimensional motorized stages. The x-ray is generated from a tungsten rotating anode x-ray source. The source grating G0 (period 22.7 μm, Au height 70 μm) is positioned about 8 cm from the emission point inside the x-ray source, the beam splitter grating G1 (period 4.36 μm, Au height 2.43 μm) which induced a phase shift of $\pi/2$ approximately at 27 keV, situates 106.9 cm from G0 behind the gantry axis, and the sample is mounted closely before G1. The analyzer grating G2 (period 5.4 μm, Au height 65 μm) is positioned at ~10 mm before the detector while the distance between G1 and G2 is 25.6 cm, corresponding to the first order integer Talbot distance of grating G1. The x-ray detector is a combination of a scintillator and a CCD camera connected via a fiber coupling charge coupled device, its pixel size is 18×18 um$^2$ and the effective receiving area is 6.84×6.84 cm$^2$.

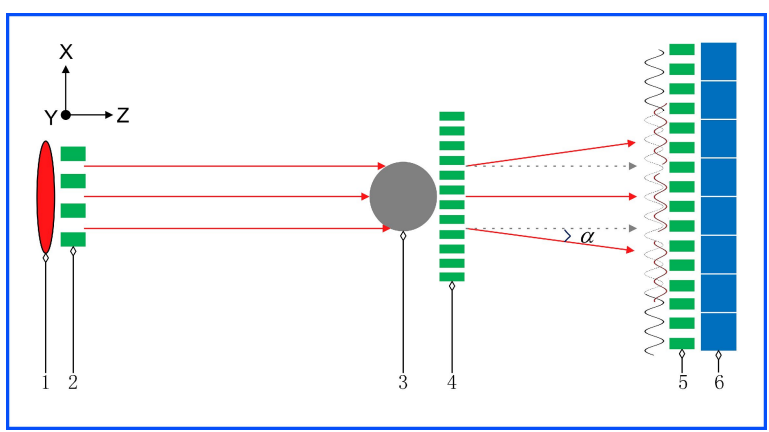

Figure 1. (Color online). Schematic of x-ray Talbot-Lau interferometer. 1. X-ray source, 2. Source grating (G0), 3. Sample, 4. Beam splitter grating (G1), 5. Analyzer grating (G2), 6. X-ray detector.

As illustrated in Fig. 1, the differential phase-contrast image information process essentially relies on the fact that the sample placed in the x-ray beam path causes slight refraction of the beam transmitted through the object, and the idea of differential phase-contrast imaging depends on locally detecting the angular deviation $\alpha$. Typically, the refraction angle can be measured by combining the moire fringe and phase-stepping technique. [28, 29] Much detailed theoretical basis of x-ray phase-contrast imaging using grating interferometer can be referred at here. [12-14, 30]

### 2.2 Image acquisition

The sample we used is a POM cylinder (its diameter is 10 mm and the length is about 90 mm), a PMMA cylinder (its diameter is 5 mm and its length is about 100 mm) and a POM cylinder (its diameter is 5 mm and length is about 80 mm). During the experiment, the x-ray generator was operated with a tube current of 45 mA and a tube accelerating voltage of 40 KV. After finely aligning the three gratings and the sample stage, 5 steps in one period of grating G2 were adopted for the phase stepping scan, and for each step, 20 s was taken to capture a raw image, then a same phase stepping scan was performed after removing the sample. Each position of grating G2 in the scan without sample was kept the same as the corresponding one in the scan with sample in the beam path by using stage of high positioning precision.

## 3. Retrieval of the refraction signal in data post processing

### 3.1 Background correction using methods arg(S/B) and arg(S)-arg(B)

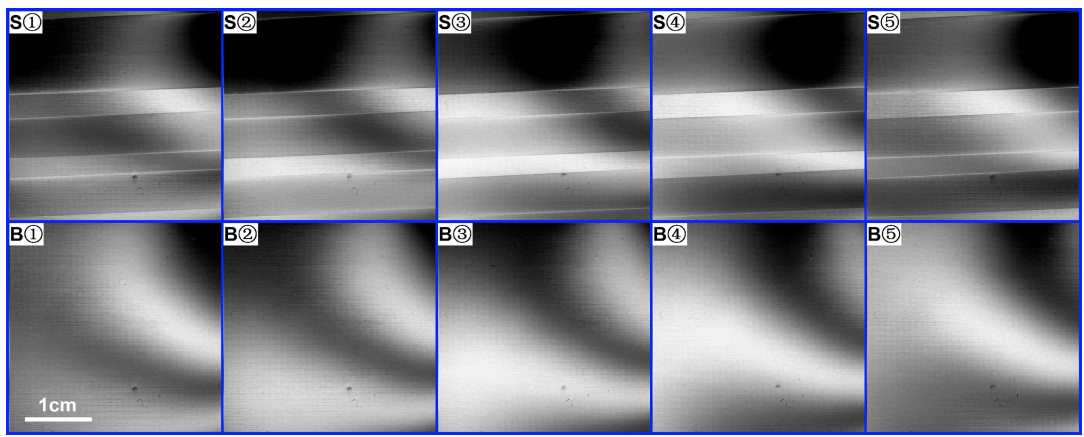

Figure 2. Raw images of the two scans. S①, S②, S③, S④, S⑤ are the data with sample and B①, B②, B③, B④, B⑤ stand for that without sample in the beam path. In the images with sample, top to bottom - POM cylinder (Φ=10mm), PMMA cylinder (Φ=5mm) and POM cylinder (Φ=5mm). All the images are windowed for optimized appearance with a linear gray scale.

Fig. 2 demonstrates raw images of the two phase stepping scans. Figs .2(S①), 2(S②), 2(S③), 2(S④) and 2(S⑤) are the 5 raw images with sample, while Figs .2(B①), (B②), 2(B③), 2(B④) and 2(B⑤) stand for those without sample (background) in the beam path. Here we like to point out that the light intensity is not uniform in the field of view of all the 5 background images because of the unideal moire fringes, as a matter of fact, it is always almost impossible to obtain a flat background with an x-ray Talbot-Lau interferometer because imperfections exist in the x-ray gratings and the alignment of gratings is not ideal.

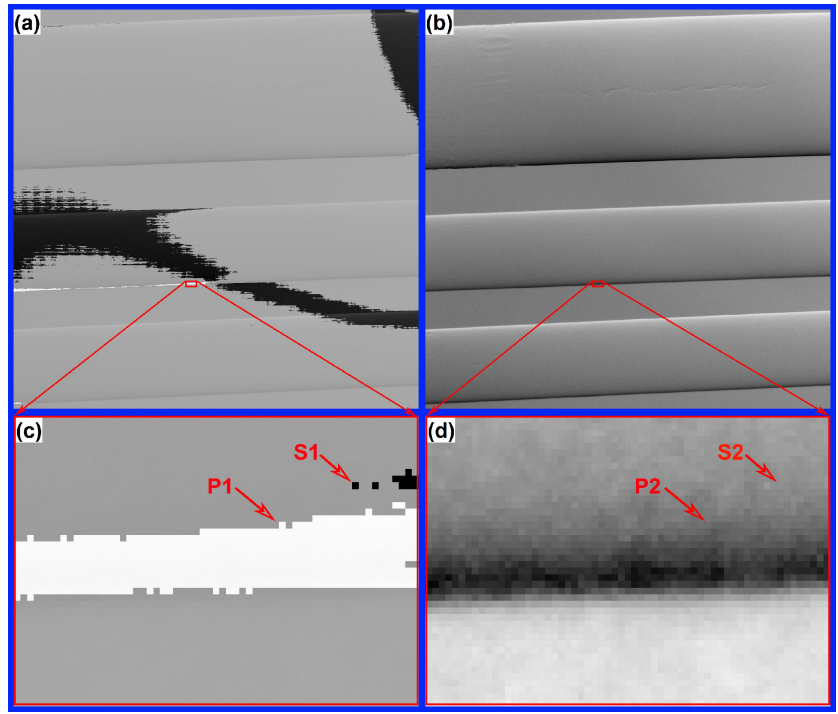

Figure 3. X-ray refraction images of the sample. The data was computed with: (a) method arg(S)-arg(B), and (b) method arg(S/B). (c) and (d) are the enlarged view of the red rectangular chosen in (a) and (b), respectively. All the images are windowed for optimized appearance with a linear gray scale.

We used methods arg(S)-arg(B) and arg(S/B) separately to compute the pure refraction image of the sample. Fig. 3(a) depicts the refraction signal yielded by method arg(S)-arg(B), while Fig. 3(b) shows that by method arg(S/B). Figs. 3(c) and 3(d) represent the enlarged view of the red rectangular extracted from the same position as shown in Figs. 3(a) and 3(b), respectively.

It was observed very clear that great differences exist between the two refraction images, for example, in Fig. 3(c) the gray values of the pixels P1 and S1 are obviously different from those of the corresponding pixels (P2 and S2) in Fig. 3(d). Here, according to our aforehand knowledge about the feature of the sample, judgment can be made that a correct refraction image was obtained by method arg(S/B) and on the contrary, the refraction angles in many parts of the image retrieved by method arg(S)-arg(B) are wrong.

**3.2 The origin of error in the refraction image retrieved by method arg(S)-arg(B)**

In the hope of figuring out the origin of error in the refraction image yielded by method arg(S)-arg(B), we traced the computational process of the refraction angle at pixel P1 as shown in Fig. 3(c) and at the corresponding pixel P2 in Fig. 3(d), which are written respectively as follows:

$$\left.\begin{aligned}\varphi_{P1} &= \frac{d}{2\pi \times z_T} \times \left\{\arg\sum_{k=0}^{4}[I_{k+1}^{s}(x,y)\times\exp(-i2\pi\frac{k}{5})] - \arg\sum_{k=0}^{4}[I_{k+1}^{b}(x,y)\times\exp(-i2\pi\frac{k}{5})]\right\} \\ &= \frac{d}{2\pi \times z_T} \times \left\{\arg[\text{-3629.9+66.9497 i}] - \arg[\text{-3899.72-499.816 i}]\right\} \\ &= \frac{d}{2\pi \times z_T} \times \{3.12315 - (-3.01412)\} \\ &= \frac{d}{2\pi \times z_T} \times 6.13727. \end{aligned}\right\} \quad (1.3)$$

$$\left.\begin{aligned}\varphi_{P2} &= \frac{d}{2\pi \times z_T} \times \arg\left\{\frac{\sum_{k=0}^{4}[I_{k+1}^{s}(x,y)\times\exp(-i2\pi\frac{k}{5})]}{\sum_{k=0}^{4}[I_{k+1}^{b}(x,y)\times\exp(-i2\pi\frac{k}{5})]}\right\} \\ &= \frac{d}{2\pi \times z_T} \times \arg\left\{\frac{\text{-3629.9+66.9497 i}}{\text{-3899.72-499.816 i}}\right\} \\ &= \frac{d}{2\pi \times z_T} \times \arg\{0.913604 - 0.134262\text{ i}\} \\ &= \frac{d}{2\pi \times z_T} \times (-0.145914). \end{aligned}\right\} \quad (1.4)$$

Similarly, we show the source of data at pixel S1 in Fig. 3(c) and at the corresponding pixel S2 in Fig. 3(d) respectively as follows:

$$\begin{aligned}
\varphi_{S1} &= \frac{d}{2\pi \times z_T} \times \left\{ \arg \sum_{k=0}^{4}[I_{k+1}^{s}(x,y) \times \exp(-i2\pi\frac{k}{5})] - \arg \sum_{k=0}^{4}[I_{k+1}^{b}(x,y) \times \exp(-i2\pi\frac{k}{5})] \right\} \\
&= \frac{d}{2\pi \times z_T} \times \{\arg[-4044.07-252.068\ i] - \arg[-4411.39+80.1258\ i]\} \\
&= \frac{d}{2\pi \times z_T} \times \{-3.07934 - 3.12343\} \\
&= \frac{d}{2\pi \times z_T} \times (-6.20277).
\end{aligned} \quad (1.5)$$

$$\begin{aligned}
\varphi_{S2} &= \frac{d}{2\pi \times z_T} \times \arg \left\{ \frac{\sum_{k=0}^{4}[I_{k+1}^{s}(x,y) \times \exp(-i2\pi\frac{k}{5})]}{\sum_{k=0}^{4}[I_{k+1}^{b}(x,y) \times \exp(-i2\pi\frac{k}{5})]} \right\} \\
&= \frac{d}{2\pi \times z_T} \times \arg\left\{ \frac{-4044.07-252.068\ i}{-4411.39+80.1258\ i} \right\} \\
&= \frac{d}{2\pi \times z_T} \times \arg\{0.915393 + 0.0737667\ i\} \\
&= \frac{d}{2\pi \times z_T} \times 0.080411.
\end{aligned} \quad (1.6)$$

The comparisons between formulae (1.3) and (1.4), between formulae (1.5) and (1.6) well confirm the differences between the refraction images retrieved by methods arg(S)-arg(B) and arg(S/B). The values of pixels P1 and S1 in Fig. 3(c) are $\frac{d}{2\pi \times z_T} \times 6.13727$ and $\frac{d}{2\pi \times z_T} \times (-6.20277)$, remember that the function range of arg(S)-arg(B) is (-2π, 2π], this is the reason why as shown in gray value image Fig. 3(c), pixel P1 features ultra-white and pixel S1 displays ultra-black. And according to our aforementioned judgment, the refraction angle at pixels P1 and S1 computed by formulae (1.3) and (1.5) are wrong.

A vivid explanation on the origin of error in the refraction image computed by method arg(S)-arg(B) can be shown in Fig. 4, the blue sinusoid curve in Fig. 4(a) stands for the shift curve of one certain pixel in the detector without sample, and the black one represents that with sample in the beam path, the lateral shift of the curve is caused by the deflection of the beam when passing through the sample, which is the fundamental principle of refraction signal retrieval in x-ray phase contrast imaging with Talbot-Lau interferometer. Here the absorption and scattering effects of the sample were ignored when generating the curve. The blue saw tooth curve in Fig. 4(b) is the phase of the pixel, possessing a one-to-one relationship with the blue sinusoid curve in Fig. 4(a) in this way, for one pixel, for example the 1$^{st}$ step of the phase stepping scan without sample is located at point A in the shift curve, and the 2$^{nd}$, 3$^{rd}$, 4$^{th}$, 5$^{th}$ steps are successively equally spaced by d/5 to the right of point A in the curve (d is period of the analyzing grating), then the retrieved phase of this pixel would be the ordinate value of point B in the saw tooth curve as shown in Fig 4(b) by formula written as:

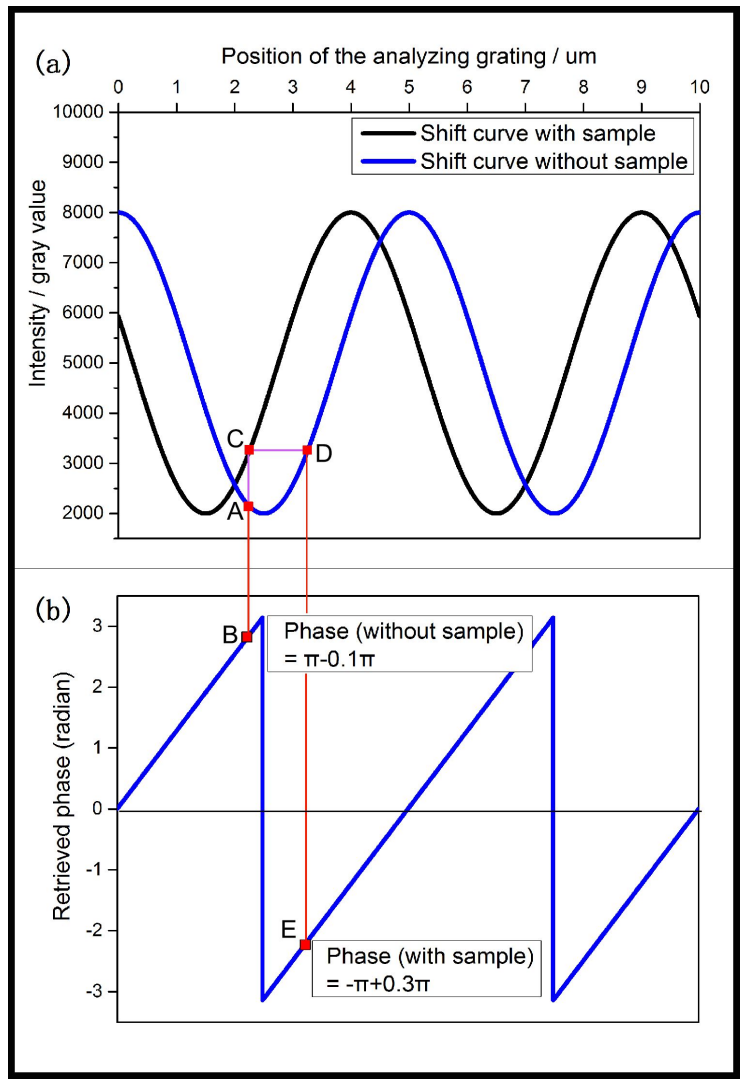

Figure 4. (Color online). The source of error in the refraction image yielded by method arg(S)-arg(B).

$$\varphi = \arg \sum_{k=0}^{4} [I_{k+1} \times \exp(-i2\pi \frac{k}{5})]. \tag{1.7}$$

where $I_{k+1}$ represents gray value of this pixel at the $(k+1)^{th}$ step of the phase stepping scan. Here we assume the phase of this pixel without sample (arg(B)) is π-0.1π, and after moving the sample inside the beam path, we make the hypothesis that the shift curve of this pixel is lateral shifted less than a period to be as shown the black sinusoid curve in Fig. 4(a).

Because each position of grating G2 in the scan with sample was kept the same as the corresponding one in the scan without sample in the beam path, the 1st step of the pixel with sample would be located at point C in the black shift curve, obviously the phase of point C in the black curve equals to that of point D in the blue shift curve. And based on the aforementioned one-to-one relationship, the phase of this pixel with sample is the ordinate value of point E as shown in Fig. 4(b), here the phase (arg(S)) is assumed to be –π+0.3π. Then the pure phase of the sample at this pixel computed by method arg(S)-arg(B) would be:

$$\arg(S) - \arg(B) = (-\pi + 0.3\pi) - (\pi - 0.1\pi) = -2\pi + 0.4\pi. \tag{1.8}$$

Clearly, this measured phase is wrong because of the discontinuous nature of the saw tooth curve as shown in Fig 4(b) (see the jumping edge between points B and E). This type of error can also be named as phase wrapping effect because the real phase $\varphi$ is:

$$\varphi = \phi + 2\pi, \tag{1.9}$$

where $\phi$ is the measured phase, error in pixel S1 can be classified into this case, while for pixel P1, the equation is:

$$\varphi = \phi - 2\pi. \tag{1.10}$$

### 3.3 Typical phase unwrapping algorithm for method arg(S)-arg(B)

To fix the error in method arg(S)-arg(B), a typical phase unwrapping solution based on the above discussion can be written as:

$$\varphi(x, y) = \frac{d}{2\pi \times z_T} \times \{[\arg(S) - \arg(B)] \bmod 2\pi\}. \tag{1.11}$$

The $[\arg(S) - \arg(B)]$ modulo $2\pi$ operation can be rewritten in the form:

$$[\arg(S) - \arg(B)] \bmod 2\pi = \arg(S) - \arg(B) + 2k\pi. \tag{1.12}$$

where $k \in \mathbb{Z}$ and forces:

$$-\pi < \arg(S) - \arg(B) + 2k\pi \leq \pi. \tag{1.13}$$

Using this phase unwrapping algorithm, correct refraction signals in pixels P1 and S1 can be obtained as:

$$\varphi_{P1}^* = \frac{d}{2\pi \times z_T} \times (6.13727 \bmod 2\pi) = \frac{d}{2\pi \times z_T} \times (-0.145915) = \varphi_{P2}, \tag{1.14}$$

$$\varphi_{S1}^* = \frac{d}{2\pi \times z_T} \times (-6.20277 \bmod 2\pi) = \frac{d}{2\pi \times z_T} \times 0.0804153 = \varphi_{S2}. \tag{1.15}$$

The refraction image of the above experimental data yielded by this phase unwrapping algorithm together with the numerical comparison would be provided later in section 4.3.

## 4 A new phase unwrapping solution for method arg(S)-arg(B)

### 4.1 Strategy of cyclic shift operation on the raw images

Our new phase unwrapping solution is based on a novel cyclic shift operation on the raw images, as shown in Fig. 5, the blue sinusoid curve in Fig. 5(a) and the blue saw tooth curve in Fig. 5(b) share the same one-to-one relationship as that in Fig. 4. Now we suppose for one pixel the phase stepping scan was performed without sample in the beam path when the analyzing grating situates successively at positions ①②③

④⑤ in the shift curve, these 5 positions are equally space by d/5 (d is the period of the analyzing grating). Then if we input the image sequence of ①②③④⑤ into formula (1.7), as described aforementioned, the retrieved phase of this pixel would be indicated as shown in Fig. 5(b) the ordinate value of point T1. Now let us cyclically shift the order of the images from ①②③④⑤ to ⑤①②③④ and calculate again the phase using formula (1.7). Clearly the new phase is identical to that of the image order of ⑤¹①②③④ as shown in Fig. 5(a) (here position ⑤¹ and ⑤ share the same feature in the sinusoid curve), and the phase of the image sequence of ⑤¹①②③④ is the ordinate value of point T2. Similarly, the phases of the image sequences of ④⑤①②③, ③④⑤①② and ②③④⑤① equal respectively to these of the image sequences of ④¹⑤¹①②③, ③¹④¹⑤¹①② and ②¹③¹④¹⑤¹①, the corresponding phases are the ordinate values of points T3, T4 and T5 as demonstrated in Fig. 5(b).

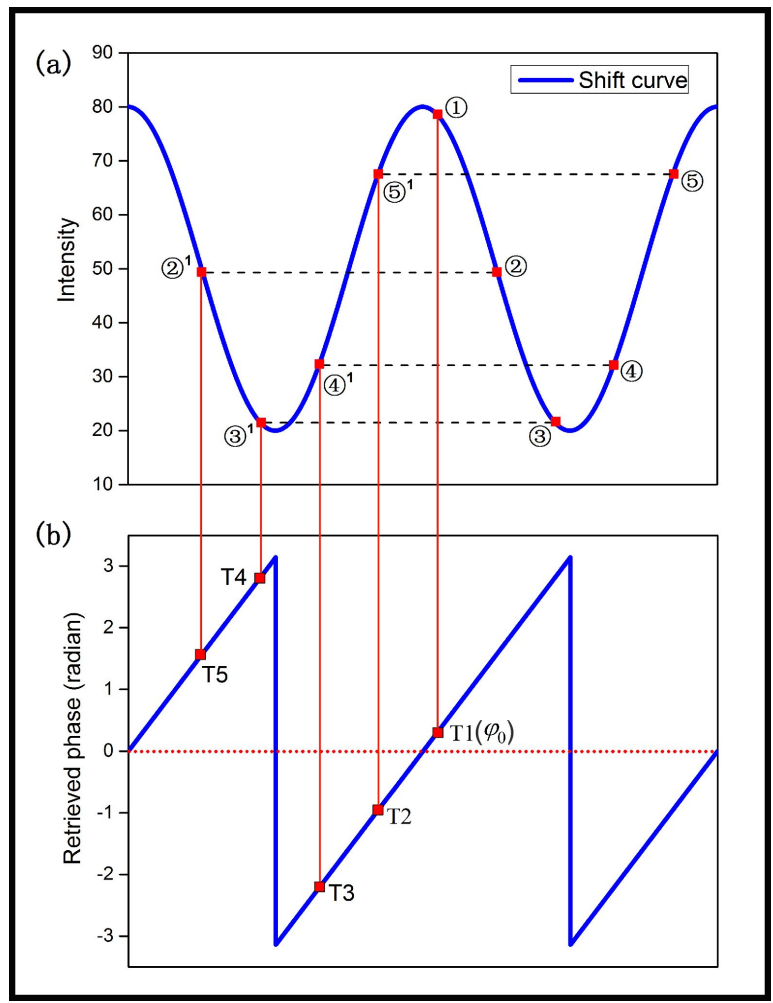

Figure 5. (Color online). Schematic of the theory of cyclic shift operation on the raw images.

In the case demonstrated in Fig. 5(b), the pixel's phase without sample in the beam path of the five image sequences can be written as:

$$\left.\begin{aligned}
&arg\{background①②③④⑤\}=\varphi_0; \\
&arg\{background⑤①②③④\}=arg\{background⑤^1①②③④\}=\varphi_0-\frac{2\pi}{5}; \\
&arg\{background④⑤①②③\}=arg\{background④^1⑤^1①②③\}=\varphi_0-\frac{4\pi}{5}; \\
&arg\{background③④⑤①②\}=arg\{background③^1④^1⑤^1①②\}=\varphi_0+\frac{4\pi}{5}; \\
&arg\{background②③④⑤①\}=arg\{background②^1③^1④^1⑤^1①\}=\varphi_0+\frac{2\pi}{5}.
\end{aligned}\right\} \quad (1.16)$$

Now after moving the sample inside the beam path, we assume the phase shift introduced by the sample $\frac{2\pi}{d}z_T\varphi$ is so tiny, for example -0.001 radian, that the contribution of the phase shift introduced by the sample would not pass over the jumping edge at points T1, T2, T3, T4 and T5, then the phases of the pixel with sample in five image orders can be expressed as:

$$\left.\begin{aligned}
&arg\{sample①②③④⑤\}=\varphi_0+\frac{2\pi}{d}z_T\varphi; \\
&arg\{sample⑤①②③④\}=arg\{sample⑤^1①②③④\}=\varphi_0+\frac{2\pi}{d}z_T\varphi-\frac{2\pi}{5}; \\
&arg\{sample④⑤①②③\}=arg\{sample④^1⑤^1①②③\}=\varphi_0+\frac{2\pi}{d}z_T\varphi-\frac{4\pi}{5}; \\
&arg\{sample③④⑤①②\}=arg\{sample③^1④^1⑤^1①②\}=\varphi_0+\frac{2\pi}{d}z_T\varphi+\frac{4\pi}{5}; \\
&arg\{sample②③④⑤①\}=arg\{sample②^1③^1④^1⑤^1①\}=\varphi_0+\frac{2\pi}{d}z_T\varphi+\frac{2\pi}{5}.
\end{aligned}\right\} \quad (1.17)$$

Than after background correction using method arg(S)-arg(B), we can get:

$$\left.\begin{aligned}
&arg\{sample①②③④⑤\}-arg\{background①②③④⑤\}=\frac{2\pi}{d}z_T\varphi; \\
&arg\{sample⑤①②③④\}-arg\{background⑤①②③④\}=\frac{2\pi}{d}z_T\varphi; \\
&arg\{sample④⑤①②③\}-arg\{background④⑤①②③\}=\frac{2\pi}{d}z_T\varphi; \\
&arg\{sample③④⑤①②\}-arg\{background③④⑤①②\}=\frac{2\pi}{d}z_T\varphi; \\
&arg\{sample②③④⑤①\}-arg\{background②③④⑤①\}=\frac{2\pi}{d}z_T\varphi.
\end{aligned}\right\} \quad (1.18)$$

Formula (1.18) shows that under the assumption that the phase shift introduced by the sample is very tiny, if the image sequence of the raw data with sample is kept the same as that of the data without sample in the beam path, the retrieved phase of the pixel by method arg(S)-arg(B) would be equal.

Fig. 6 demonstrates the computational process using method arg(S)-arg(B) in five image sequences of the aforementioned experimental data, Figs. 6(A1), 6(B1) and 6(C1) show successively the phase image with sample, the phase image without sample and the pure phase image of the sample when the image order is ①②③④⑤. Similarly, Figs. 6(A2), 6(B2), 6(C2), Figs. 6(A3), 6(B3), 6(C3), Figs. 6(A4), 6(B4), 6(C4) and Figs. 6(A5), 6(B5), 6(C5) represent respectively those when the image sequence are ⑤①②③④, ④⑤①②③, ③④⑤①② and ②③④⑤①.

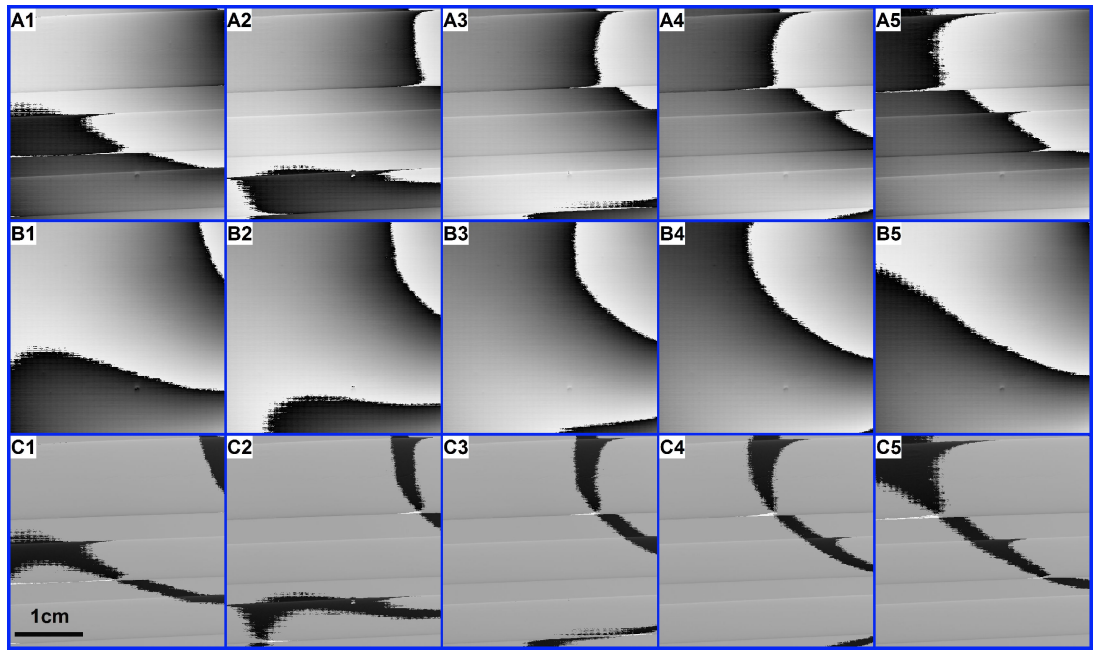

Figure 6. A1, B1 and C1 show the computational process of arg(S)-arg(B) when the image sequence is ①②③④⑤, A1: phase image with sample, B1: phase image without sample, C1: pure phase image of the sample. A2/B2/C2, A3/B3/C3, A4/B4/C4, A5/B5/C5 show respectively those when the image sequence are ⑤①②③④, ④⑤①②③, ③④⑤①② and ②③④⑤①.
All the images are windowed for optimized appearance with a linear gray scale.

Comparison among the 5 pure phase images of the sample as shown in Figs. 6(C1), 6(C2), 6(C3), 6(C4), 6(C5) shows that in contrast with the theoretical analysis in formula (1.18), the retrieved phase images of the five image sequences are totally different, meanwhile, it was observed that the areas where errors take place is always moving in the 5 phase images, the reason is that the phase shift introduced by the sample in this experiment is not so tiny as we assumed in formula (1.17) that the contribution of the sample' phase shift would pass over the jumping edge, and according to our aforementioned analysis, the error is always happening when the initial phase is located near the jumping edge of the saw tooth curve as shown in Fig. 4(b), in another way we can say that error usually occurs at the place where the light intensity in the first image of the data situates around the minimum value as demonstrated in Fig. 4(a) (this was confirmed as we can see in Fig. 6(C1) the traces of error coincide to some range the dark areas in Fig. 2(B①)). If we change the image sequence of the raw data, the dark parts in the first image would move because the moire fringe is shifting when the position of the analyzing grating is changed in phase stepping scan, and therefore the areas where errors take place would be moving in the 5 phase images.

The phenomenon that the trace of error is always moving in the 5 phase images implies that for each pixel, there would surely be at least one pure phase image without introducing error. For example, see the upper right part of the phase image, error appeared only in Figs. 6(C1) and 6(C2), while correct results were obtained in Figs. 6(C3), 6(C4) and 6(C5). A judgment criteria on which phase image of the 5 image sequences possess the least possibility of yielding error for a certain pixel can be made, see Fig. 5(b), the criteria can be achieved that error has the least possibility of emerging in the image sequence whose initial phase at this pixel without sample is located as far from the jumping edge as possible, it is identical to that the initial phase situates as near the red dotted line as possible. This criteria works because for the pixel, it would be the safest condition where the subsequent phase with the contribution of the sample' phase shift would not cross the jumping edge if we beforehand do not know the direction of the phase shift introduced by the sample, remember that the jumping edge is the ultimate source of error in method arg(S)-arg(B).

**4.2 Algorithm of the new phase unwrapping solution**

S1. Calculate the phase images with and without sample respectively by formula (1.7) in 5 image sequences of the raw data as follows:

$$\begin{aligned}
\theta_1^b(x,y) &= \arg\{\text{background ①②③④⑤}\}, & \theta_1^s(x,y) &= \arg\{\text{sample ①②③④⑤}\}; \\
\theta_2^b(x,y) &= \arg\{\text{background ⑤①②③④}\}, & \theta_2^s(x,y) &= \arg\{\text{sample ⑤①②③④}\}; \\
\theta_3^b(x,y) &= \arg\{\text{background ④⑤①②③}\}, & \theta_3^s(x,y) &= \arg\{\text{sample ④⑤①②③}\}; \\
\theta_4^b(x,y) &= \arg\{\text{background ③④⑤①②}\}, & \theta_4^s(x,y) &= \arg\{\text{sample ③④⑤①②}\}; \\
\theta_5^b(x,y) &= \arg\{\text{background ②③④⑤①}\}, & \theta_5^s(x,y) &= \arg\{\text{sample ②③④⑤①}\}.
\end{aligned}$$ (1.19)

S2. For pixel (1, 1), index the minimum absolute value in the 5 phase images without sample as:

$$k = \text{Minimum index}\left[\left|\theta_1^b(1,1)\right|, \left|\theta_2^b(1,1)\right|, \left|\theta_3^b(1,1)\right|, \left|\theta_4^b(1,1)\right|, \left|\theta_5^b(1,1)\right|\right] \quad 1 \le k \le 5.$$ (1.20)

S3: Compute the refraction angle of pixel (1, 1) as:

$$\varphi(1,1) = \frac{d}{2\pi \times z_T} \times \left[\theta_k^s(1,1) - \theta_k^b(1,1)\right].$$ (1.21)

Step 4: For all the other pixels, repeat S2 and S3, and the refraction image $\varphi(x,y)$ is obtained.

## 4.3 Performance of the new phase unwrapping solution

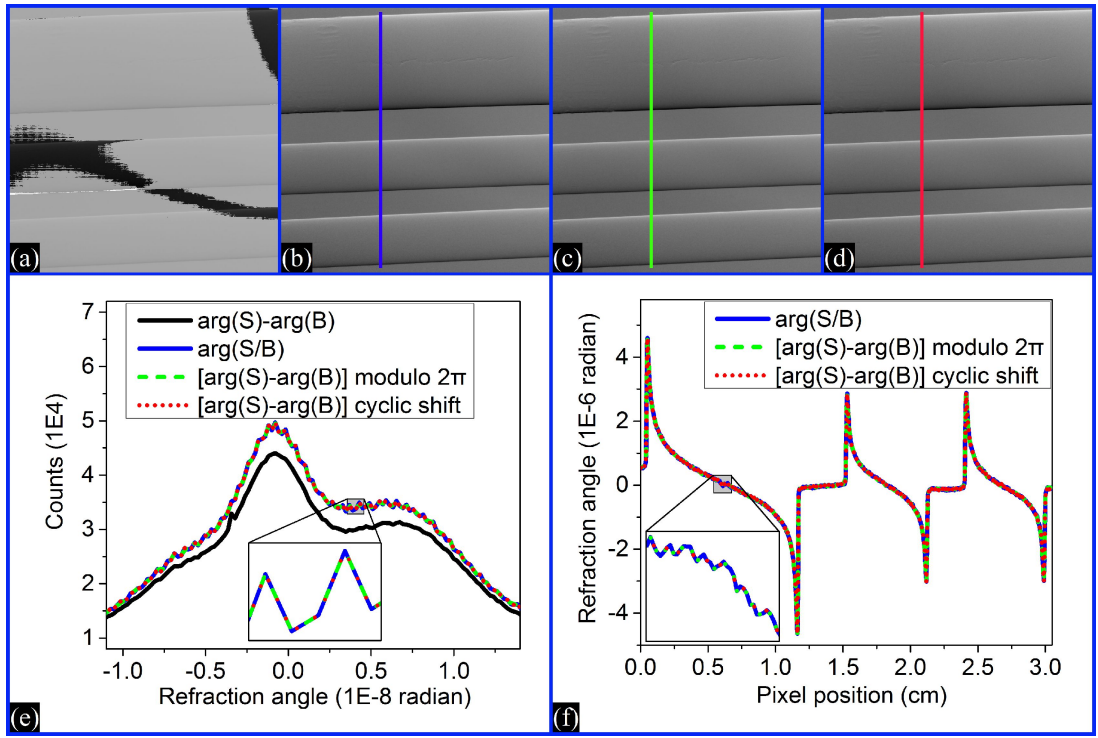

Figure 7. (Color online). Refraction images of the sample after background correction with (a): method arg(S)-arg(B), (b) method arg(S/B), (c) phase unwrapping solution [arg(S)-arg(B)] modulo $2\pi$, and (d) the new phase unwrapping algorithm. (e) are histograms of the four images, and (f) represent the refraction angles of the cross sections chosen as shown in (b), (c) and (d) respectively. All the images are windowed for optimized appearance with a linear gray scale.

The first case we took for evaluating the new phase unwrapping solution is the aforementioned experimental data. Figs. 7(a), 7(b), 7(c) and 7(d) show respectively the retrieved phase image of the sample by method arg(S)-arg(B), method arg(S/B), phase unwrapping solution [arg(S)-arg(B)] modulo $2\pi$ and the new phase unwrapping algorithm (we name it as [arg(S)-arg(B)] cyclic shift). In Fig. 7(e), the black curve, the blue curve, the green dashed curve and the red dotted one depict respectively histograms of the four images, and the range of refraction angle displayed is from -1.1E-8 to 1.4E-8 radian. Fig. 7(f) is the refraction angles of the cross sections chosen as shown the lines in Figs. 7 (b), (c) and (d) respectively, the blue curve stands for the data yield by method arg(S/B), while the green dashed one and the red dotted one shows those of the two phase unwrapping solutions. Here 10 pixels in the horizontal direction were averaged to generate the profiles, in the hope of decreasing the noise.

Fig. 8 shows performance of the new phase wrapping solution when dealing with another set of data, the sample is a chicken bone, and raw images were acquired in a same manner as the previous one. Fig. 8(a) represents the refraction image obtained by method arg(S)-arg(B), and Fig. 8(b) shows that by method arg(S/B), while Figs. 8(c) and 8(d) depict those using the two phase unwrapping solutions. Fig. 8(e) provide histograms of the four images, we chose the refraction angle from -11.5E-7 to -5E-7 radian when generating the histogram, the black curve, the blue curve, the green dashed curve and the red dotted one show respectively histograms of the four images. Fig. 8(f) are the refraction angles of the cross sections chosen as shown in Figs. 8(b), 8(c) and 8(d) respectively, the blue curve stands for the profile of method arg(S/B), while the green dashed curve and the red dotted one are those of the two phase unwrapping algorithms.

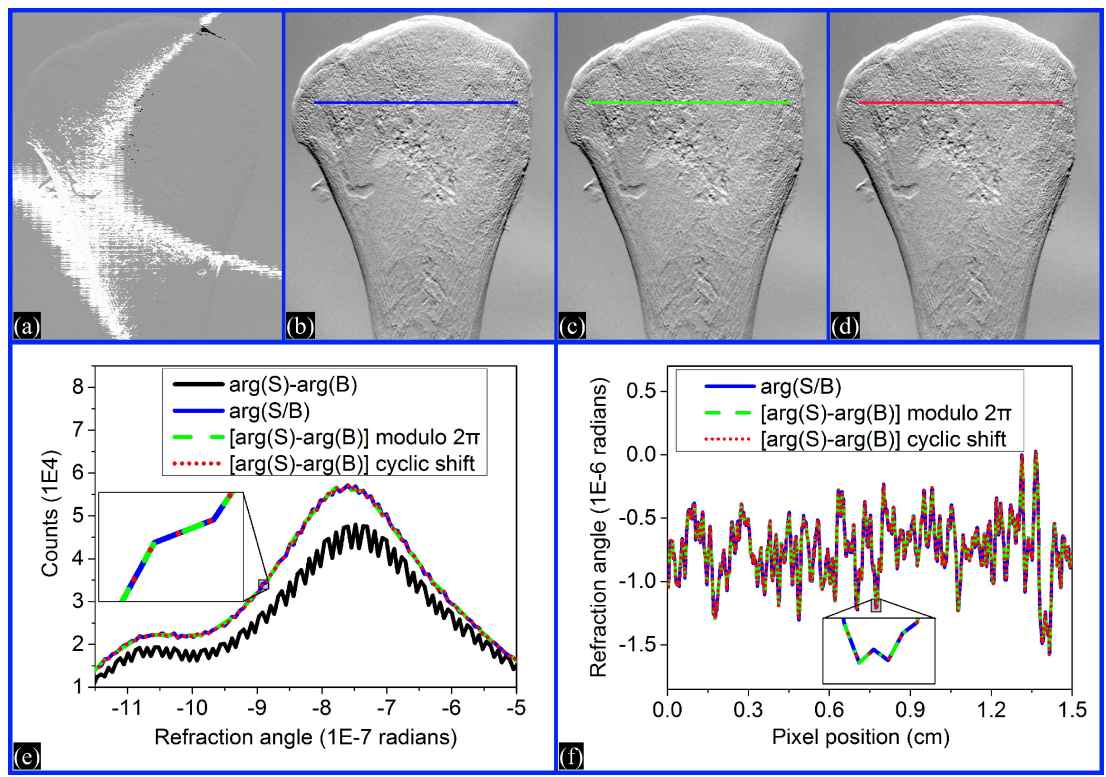

Figure 8. (Color online). Refraction images of the sample after background correction with: methods (a) arg(S)-arg(B) and (b) arg(S/B); phase unwrapping solutions (c) [arg(S)-arg(B)] modulo 2π and (d) [arg(S)-arg(B)] cyclic shift. (e) are histograms of the four images, and (f) shows the refraction angles of the cross sections chosen as shown in (b), (c) and (d) respectively. All the images are windowed for optimized appearance with a linear gray scale.

## 5  Discussion

From Figs. 7(c), 7(d), 8(c) and 8(d), it was observed very clear that using both the two introduced phase unwrapping solutions, the phase wrapping phenomenon in the refraction images by method arg(S)-arg(B) as shown in Figs. 7(a) and 8(a) has been totally eliminated and also we can see that there is almost no visual difference among the images yielded by the two phase unwrapping solutions and the images computed by method arg(S/B). Numerical comparisons as shown in Figs. 7(e), 8(e) the histograms and in Figs. 7(f), 8(f) the section profiles further confirm us that both the two phase unwrapping strategies could successfully yield exactly the same refraction signal as method arg(S/B), see the enlarged view in Figs. 7(e), 7(f), 8(e) and

8(f).

It should be pointed out that the first phase unwrapping solution ([arg(S)-arg(B)] modulo 2π) shares the same feature with method arg(S/B), the effective phase measuring range of both is fixed to be (-π, π], and phase wrapping effect would not take place if the desired phase situates within this range. On the contrary, there is some differences between solution [arg(S)-arg(B)] cyclic shift and method arg(S/B), in the aforementioned data post processing, we tuned the initial phase of every pixel without sample to position T1 as shown in Fig. 5(b), and then this strategy could measure without error the phase in the range of $(-\varphi_0 - \pi, -\varphi_0 + \pi]$ (we assume $\varphi_0$ is a positive number for example $\frac{\pi}{10}$). If we move the initial phase of every pixel to point T3, then the phase in the range of $\left(-\varphi_0 - \frac{\pi}{5}, -\varphi_0 + \frac{9\pi}{5}\right]$ can be measured by this phase unwrapping algorithm without error, remember that the effective phase measuring range of method arg(S/B) is fixed to be (-π, π], therefore the nature that the effective phase measuring range of the new phase unwrapping solution could be tuned in some degree would embody its superiority over method arg(S/B) in certain cases.

For an easier implementation of the new phase unwrapping solution in computer, the first step S1 in section 4.2 can be replaced as follows without the need of cyclically shifting the sequence of the raw images.

$$\left. \begin{array}{l} \theta_1^b(x,y)=\arg\sum_{k=0}^{4}[I_{k+1}^b \times \exp(-i2\pi\frac{k}{5})], \ \theta_1^s(x,y)=\arg\sum_{k=0}^{4}[I_{k+1}^s \times \exp(-i2\pi\frac{k}{5})]; \\ \theta_2^b(x,y)=\arg\sum_{k=1}^{5}[I_k^b \times \exp(-i2\pi\frac{k}{5})], \ \theta_2^s(x,y)=\arg\sum_{k=1}^{5}[I_k^s \times \exp(-i2\pi\frac{k}{5})]; \\ \theta_3^b(x,y)=\arg\sum_{k=2}^{6}[I_{k-1}^b \times \exp(-i2\pi\frac{k}{5})], \ \theta_3^s(x,y)=\arg\sum_{k=2}^{6}[I_{k-1}^s \times \exp(-i2\pi\frac{k}{5})]; \\ \theta_4^b(x,y)=\arg\sum_{k=3}^{7}[I_{k-2}^b \times \exp(-i2\pi\frac{k}{5})], \ \theta_4^s(x,y)=\arg\sum_{k=3}^{7}[I_{k-2}^s \times \exp(-i2\pi\frac{k}{5})]; \\ \theta_5^b(x,y)=\arg\sum_{k=4}^{8}[I_{k-3}^b \times \exp(-i2\pi\frac{k}{5})], \ \theta_5^s(x,y)=\arg\sum_{k=4}^{8}[I_{k-3}^s \times \exp(-i2\pi\frac{k}{5})]. \end{array} \right\} \quad (1.22)$$

This numerical computing is identical to formula (1.19) because:

$$\theta_2(x,y) = \left[\underbrace{\arg\sum_{k=0}^{4}[I_{k+1}\times\exp(-i2\pi\frac{k}{5})]}_{\text{sequence}=⑤①②③④}\right] = \left[\underbrace{\arg\sum_{k=1}^{5}[I_k\times\exp(-i2\pi\frac{k}{5})]}_{\text{sequence}=①②③④⑤}\right];$$

$$\theta_3(x,y) = \left[\underbrace{\arg\sum_{k=0}^{4}[I_{k+1}\times\exp(-i2\pi\frac{k}{5})]}_{\text{sequence}=④⑤①②③}\right] = \left[\underbrace{\arg\sum_{k=2}^{6}[I_{k-1}\times\exp(-i2\pi\frac{k}{5})]}_{\text{sequence}=①②③④⑤}\right];$$

$$\theta_4(x,y) = \left[\underbrace{\arg\sum_{k=0}^{4}[I_{k+1}\times\exp(-i2\pi\frac{k}{5})]}_{\text{sequence}=③④⑤①②}\right] = \left[\underbrace{\arg\sum_{k=3}^{7}[I_{k-2}\times\exp(-i2\pi\frac{k}{5})]}_{\text{sequence}=①②③④⑤}\right];$$

$$\theta_5(x,y) = \left[\underbrace{\arg\sum_{k=0}^{4}[I_{k+1}\times\exp(-i2\pi\frac{k}{5})]}_{\text{sequence}=②③④⑤①}\right] = \left[\underbrace{\arg\sum_{k=4}^{8}[I_{k-3}\times\exp(-i2\pi\frac{k}{5})]}_{\text{sequence}=①②③④⑤}\right].$$

(1.23)

Another algorithm sharing the same key point of the new phase unwrapping solution can be written as follows, in the hope of reducing the memory consumed in computer.

S1: For pixel (1, 1), index the maximum value of light intensity in the 5 raw images without sample as:

$$n = \text{Maximum index}\left[I_1^b(1,1), I_2^b(1,1), I_3^b(1,1), I_4^b(1,1), I_5^b(1,1)\right] \qquad 1 \leq n \leq 5. \qquad (1.24)$$

S2: Compute the refraction angle of pixel (1, 1) as:

$$\varphi(1,1) = \frac{d}{2\pi \times z_T} \times \left[\arg\sum_{k=1-n}^{5-n}[I_{k+n}^s \times \exp(-i2\pi\frac{k}{5})] - \arg\sum_{k=1-n}^{5-n}[I_{k+n}^b \times \exp(-i2\pi\frac{k}{5})]\right]. \qquad (1.25)$$

S3: For all the other pixels, repeat S1 and S2, and the refraction image $\varphi(x,y)$ is obtained.

This algorithm is identical to the solution described in section 4.2, as shown in Fig. 5(a), if the light intensity of the first step in the image sequence is the maximum, the initial phase would then situates as far from the jumping edge as possible. Advantage of this algorithm is that the phase was computed only once and the memory consumed in computer would thus be greatly reduced, remember that in the previous two algorithms, the phase has to be computed for 5 times and the personal computer has to allocate much more memory for the storage of the 10 phase images.

Finally, we want to point out that the new phase unwrapping solution can also get used in other techniques, for example in x-ray phase contrast imaging with a crystal x-ray interferometer, [4-6] and in phase measurement technique with visible light. [29]

# 6 Conclusion

In this study, we talked about two background correction methods in x-ray phase contrast imaging with Talbot-Lau interferometer.

(i) Method arg(S)-arg(B). Its effective phase measuring range is (-2π, 2π], but phase wrapping phenomenon usually happens when the pure phase introduced by the sample is very small.

(ii) Method arg(S/B). The effective phase measuring range is (-π, π], and phase wrapping effect wouldn't exist if the desired phase situates inside this range.

Meanwhile, we introduced two phase unwrapping solutions to fix the errors in background correction method arg(S)-arg(B).

(i) Algorithm [arg(S)-arg(B)] modulo 2π. This solution shares exactly the same feature with method arg(S/B).

(ii) Algorithm [arg(S)-arg(B)] cyclic shift. With this phase unwrapping solution, phase wrapping effect wouldn't take place when the phase to measure is small. And compared with method arg(S/B), superiority of this algorithms is that its effective phase measuring range could be tuned in some degree for example to (-π+3, π+3], thus it would find potential advantage under certain case because the effective phase measuring range of method arg(S/B) is fixed to be (-π, π].

**Acknowledgements**

The authors are grateful to Murakami Gaku, Wataru Abe, Taiki Umemoto and Kosuke Kato (*Institute of Multidisciplinary Research for Advanced Materials, Tohoku University, 2-1-1Katahira, Aoba-ku, Sendai, Miyagi 980-8577, Japan.*) for their kind help while conducting the experiments.

# References

[1] Lewis R 2004 *Phys. Med, Biol.* **49** 3573

[2] Momose A 2005 *Jpn. J. Appl. Phys.* **44** 6355

[3] Zhou S A and Brahme A 2008 *Phys. Medica* **24** 129

[4] Bonse U and Hart M 1965 *Appl. Phys. Lett.* **6** 155

[5] Momose A 1995 *Nucl. Instrum. Meth. A* **352** 622

[6] Momose A, Takeda T, Itai Y and Hirano K 1996 *Nat. Med.* **2** 596

[7] Wilkins S, Gureyev T, Gao D, Pogany A and Stevenson A 1996 *Nature* **384** 335

[8] Nugent K, Gureyev T, Cookson D, Paganin D and Barnea Z 1996 *Phys. Rev. Lett.* **77** 2961

[9] Davis T, Gao D, Gureyev T, Stevenson A and Wilkins S 1995 *Nature* **373** 595

[10] Chapman D, Thomlinson W, Johnston R, Washburn D, Pisano E, Gmür N, Zhong Z, Menk R, Arfelli F and Sayers D 1997 *Phys. Med, Biol.* **42** 2015

[11] David C, Nöhammer B, Solak H H and Ziegler E 2002 *Appl. Phys. Lett.* **81** 3287

[12] Momose A, Kawamoto S, Koyama I, Hamaishi Y, Takai K and Suzuki Y 2003 *Jpn. J. Appl. Phys.* **42** L866

[13] Momose A, Yashiro W, Takeda Y, Suzuki Y and Hattori T 2006 *Jpn. J. Appl. Phys.* **45** 5254

[14] Pfeiffer F, Weitkamp T, Bunk O and David C 2006 *Nat.Phys.* **2** 258

[15] Pfeiffer F, Kottler C, Bunk O and David C 2007 *Phys. Rev. Lett.* **98** 108105

[16] Pfeiffer F, Bech M, Bunk O, Kraft P, Eikenberry E F, Broennimann C, Gruenzweig C and David C 2008 *Nat. Mater.* **7** 134

[17] Stutman D, Beck T J, Carrino J A and Bingham C O 2011 *Phys. Med, Biol.* **56** 5697

[18] Zanette I, Weitkamp T, Le Duc G and Pfeiffer F 2013 *RSC Adv.* **3** 19816

[19] Bech M, Tapfer A, Velroyen A, Yaroshenko A, Pauwels B, Hostens J, Bruyndonckx P, Sasov A and Pfeiffer F 2013 *Sci. Rep.* **3**

[20] Tanaka J, Nagashima M, Kido K, Hoshino Y, Kiyohara J, Makifuchi C, Nishino S, Nagatsuka S and Momose A 2013 *Z. Med. Phy.* **23** 222

[21] Momose A, Yashiro W, Kido K, Kiyohara J, Makifuchi C, Ito T, Nagatsuka S, Honda C, Noda D and Hattori T 2014 *Philos. T. R. Soc. A* **372** 20130023

[22] Zanette I 2010 Interférométrie X à réseaux pour l'imagerie et l'analyse de front d'ondes au synchrotron (Ph.D dissertation) (France: Université de Grenoble).

[23] Itoh K 1982 *Appl. Opt.* **21** 2470

[24] Judge T R and Bryanston C P 1994 *Opt. Laser. Eng.* **21** 199

[25] Haas W, Bech M, Bartl P, Bayer F, Ritter A, Weber T, Pelzer G, Willner M, Achterhold K and Durst J 2011 *Proc. of SPIE* **7962** 79624R-1

[26] Jerjen I, Revol V, Schuetz P, Kottler C, Kaufmann R, Luethi T, Jefimovs K, Urban C and Sennhauser U 2011 *Opt. Express* **19** 13604

[27] Epple F, Potdevin G, Thibault P, Ehn S, Herzen J, Hipp A, Beckmann F and Pfeiffer F 2013 *Opt. Express* **21** 29101

[28] Bruning J, Herriott D R, Gallagher J, Rosenfeld D, White A and Brangaccio D 1974 *Appl. Opt.* **13** 2693

[29] Creath K 1988 *Prog. optics* **26** 349

[30] Weitkamp T, Diaz A, David C, Pfeiffer F, Stampanoni M, Cloetens P and Ziegler E 2005 *Opt. Express* **13** 6296